\documentclass[12pt]{article}
\usepackage{graphicx}
\usepackage{subfigure}
\newcommand{\beq}{\begin{equation}}
\newcommand{\eeq}{\end{equation}}
\newcommand{\bea}{\begin{eqnarray}}
\newcommand{\eea}{\end{eqnarray}}

\renewcommand{\a}{\alpha}

\newcommand{\overlrarrow}[1]{\vbox{\ialign{##\cr\cr
                  \leftrightarrowfill\crcr\noalign{\kern-1pt\nointerlineskip}
                  $\hfil\displaystyle{#1}\hfil$\crcr}}}

\begin{document}
\begin{titlepage}
\begin{flushleft}
       \hfill                      {\tt hep-th/1010.4396}\\
       \hfill                       FIT HE - 10-02 \\
       \hfill                       KYUSHU-HET 126 \\
\end{flushleft}
\vspace*{3mm}
\begin{center}
{\bf\LARGE  Accelerated Quark and\vspace*{3mm}\\
Holography for Confining Gauge theory\vspace*{3mm}
}

\vspace*{5mm}
\vspace*{2mm}
\vspace*{5mm}
{\large Kazuo Ghoroku${}^{a}$\footnote[1]{\tt gouroku@dontaku.fit.ac.jp},
Masafumi Ishihara${}^{b}$\footnote[2]{\tt masafumi.ishihara@gmail.com},
\\Kouki Kubo${}^{c}$\footnote[3]{\tt kkubo@higgs.phys.kyushu-u.ac.jp}, Tomoki Taminato${}^{c}$\footnote[4]{\tt taminato@higgs.phys.kyushu-u.ac.jp}
}\\

\vspace*{4mm}
{${}^{a}$Fukuoka Institute of Technology, Wajiro, 
Higashi-ku} \\
{
Fukuoka 811-0295, Japan\\}
{
${}^{b}$Department of Electrophysics, National Chiao-Tung University, Hsinchu, Taiwan, R.O.C.}\\
{
}
{
${}^{c}$Department of Physics, Kyushu University, Hakozaki,
Higashi-ku}\\
{
Fukuoka 812-8581, Japan\\}
\end{center}

\begin{abstract}
We show a constantly accelerated quark as a string 
solution of the Nambu-Goto action, which is embedded in the bulk
background dual to the $\cal{N}$ $=2$ 
supersymmetric confining Yang-Mills theory. The 
induced metric of the world sheet for this string solution
has an event horizon specified by the fifth coordinate.
By an extended Rindler transformation proposed by Xiao, we move
to the comoving frame of the accelerated quark-string. Then we find
that this horizon 
is transferred to the event horizon of the bulk and
the causal part of the 
accelerated quark 
is transformed to a static
free-quark in the Rindler coordinate.
As a result, 
the confinement of the Minkowski vacuum
is lost in the Rindler vacuum.
This point is assured also by studying the potential between the
quark and anti-quark.
However, the remnants of the original
confining force are seen in various thermal quantities. 
We also discuss the consistency of our results and the claim 
that the Green's functions will not be changed by the Rindler transformation.
\end{abstract}

\end{titlepage}

\section{Introduction}

The observer, who is accelerated with a constant acceleration, $a$,
in the Minkowski space-time would observe thermal bath of particles 
with the temperature $a/2\pi$ (the Rindler temperature)
\cite{Unruh:1976db}-\cite{Deser:1998xb}. 
This phenomenon is known as the Unruh effect 
(see the review for example \cite{Crispino:2007eb}). 
Recently, similar
situation has been studied for the $\cal {N}$ $=4$ supersymmetric Yang-Mills theory 
in the context of the holography \cite{Paredes:2008cr,Xiao:2008nr,Hirayama:2010xi,Caceres:2010rm}.
In the approach of \cite{Xiao:2008nr,Hirayama:2010xi}, 
an accelerated quark has been introduced as string solutions
of the Nambu-Goto action which is embedded in the $AdS_5$ background dual
to the $\cal {N}$ $=4$ supersymmetric Yang-Mills theory. 
The solution in this background
has been found by Xiao \cite{Xiao:2008nr}, and one
finds an
event horizon in the induced metric (in its world sheet) of this string configuration.
The position of this horizon is specified by the fifth coordinate of the bulk.

Xiao has proposed an extended form of Rindler transformation (ERT) to move to
a comoving frame of the accelerated quark. Performing this ERT, the 
event horizon appears in the bulk.
Thus the theory dual to the ERT
transformed geometry is considered as a finite temperature
Yang-Mills theory. The temperature is given by the Rindler temperature $a/2\pi(=T_R)$.
At the same time, the position of the bulk horizon
can be put at the same fifth-coordinate point with the one of the world sheet horizon of the
accelerated string. As a result, in the new coordinate, one finds a static string 
which connects the boundary and the event horizon
of the bulk.

This is nothing but the free quark string-configuration in the
Rindler vacuum. Since 
the theory dual to the $AdS_5$ is in the deconfinement phase, there is also a free quark
in the Minkowski vacuum at zero temperature.
However we should notice that the free quark in the one 
vacuum is not the same with the one in the other vacuum, 
because the static free-quark in the Minkowski vacuum cannot be transformed to the
one of the Rindler vacuum by ERT, and vice versa. 
Anyway, 
in both theories dual to $AdS_5$
and to the one transformed by ERT, 
the quarks are not confined and
chiral symmetry is not broken. So the confinement-deconfinement transition
or chiral symmetry
restoration has not been discussed as the thermal effect in the Rindler
vacuum.
Thus, it is remained as an important point
to perform this analysis for the gauge theory
in the confinement phase and also for the theory with the chiral symmetry
broken in the Minkowski vacuum. Up to now, there is no such attempt.

Here we consider a confining Yang-Mills theory in the Minkowski vacuum in order
to examine the property of its Rindler vacuum, which
is obtained by performing the ERT. 
As a concrete model, 
we consider a supersymmetric background solution of type IIB
theory. This background is dual to ${\cal N}=2$ supersymmetric Yang-Mills theory,
and the quark is confined in this theory\cite{Kehagias:1999iy}-\cite{Ghoroku:2004sp}.
In other words, it is impossible to find a free quark string solution
for the Nambu-Goto action embedded in this background.

Then we firstly solve
the equation of motion for the Nambu-Goto action to find a constantly accelerated
quark-string solution, which has a similar functional form to the Xiao's one.
Actually, we could find such a
solution. 
Then the original coordinate with the Minkowski vacuum
is transformed to the comoving coordinate of the accelerated string solution
by the ERT given by Xiao. 
After performing this transformation,
we could find the free-quark string configuration in the Rindler vacuum
as in the case of $AdS_5$.
This implies that this Rindler vacuum is in the quark deconfinement phase.
However, we should again notice that the quark in the Rindler vacuum is not the 
one in the Minkowski vacuum. Then this phase change between the Minkowski and 
Rindler vacuum cannot be interpreted as the phase transition, which is seen in the 
usual finite temperature theory. In the latter case, the quark is 
considered to be common in both phases.

In the new coordinate vacuum, the dual theory can be
regarded as the thermal Yang-Mills theory
with the Rindler temperature. Its thermal 
properties are then 
examined furthermore, and we could assure that the confinement has been lost 
at any value of the finite Rindler temperature. So, there is no
critical temperature in this case. On the other hand,
some remnants of the confining force 
are seen in various quantities. The situation is similar to the case
of the finite temperature theory dual to the $AdS_5$-Schwartzschild background.

One may consider that
our result seems to be inconsistent with the statement of \cite{Unruh:1983ac}.
There is a claim that
the quark deconfinement is not expected in the Rindler
vacuum. This is based on the equivalence of the vacuum expectation value of
Green's functions.
However, it can be understood that our calculations and the results derived from them
are not contradicting with the claim
given in \cite{Unruh:1983ac} based on the Green's functions since 
our results are derived from the Wilson-loop calculation in each vacuum
for the corresponding quarks, which cannot be related by the coordinate transformation
as mentioned above. 
This point is explained more
in the section 5.

\vspace{.2cm}
In section 2, we give the setting of our model for the supersymmetric  
confining Yang-Mills theory. In section 3, the accelerated string solutions
for the supersymmetric theory are given, and then the effect
of the confining force is examined by comparing with the solution given
for $AdS_5$ background. In section 4, new coordinate is
given by the ERT, and  
we find the same Rindler temperature with the one given for $AdS_5$.
The thermal effects are studied to see that the confinement is lost
in the Rindler vacuum.
However, the remnant of the confining
force has been observed in the Wilson-loop calculation and the drag
force.
In section 5, a brief comment
related to the work of 4D field theory is given.
The summary and discussions are given in the final section. 

\section{D3 model for confining Yang-Mills theory}

We consider 10D IIB model retaining the dilaton
$\Phi$, axion $\chi$ and self-dual five form field strength $F_{(5)}$.
Under the Freund-Rubin
ansatz for $F_{(5)}$, 
$F_{\mu_1\cdots\mu_5}=-\sqrt{\Lambda}/2~\epsilon_{\mu_1\cdots\mu_5}$ 
\cite{Kehagias:1999iy,Liu:1999fc}, and for the 10D metric as $M_5\times S^5$ or
$ds^2=g_{MN}dx^Mdx^N+g_{ij}dx^idx^j$, we find the solution.
The five dimensional $M_5$ part of the
solution is obtained by solving the following reduced 5d action,
\beq\label{10d-action}
 S={1\over 2\kappa^2}\int d^5x\sqrt{-g}\left(R+3\Lambda-
{1\over 2}(\partial \Phi)^2+{1\over 2}e^{2\Phi}(\partial \chi)^2
\right), \label{5d-action}
\eeq
which is written 
in the string frame and taking $\alpha'=g_s=1$. 

\vspace{.3cm}
The solution is obtained under the ansatz,
\beq
\chi=-e^{-\Phi}+\chi_0 \ ,
\label{super}
\eeq
which is necessary to obtain supersymmetric solutions. 
And the solution is expressed as
$$ 
ds^2_{10}=G_{MN}dX^{M}dX^{N} ~~~~~~~~~~~~~~\qquad
$$ 
\beq\label{background}
=e^{\Phi/2}
\left\{
{r^2 \over R^2}A^2(r)\left(-dt^2+(dx^i)^2\right)+
\frac{R^2}{r^2} dr^2+R^2 d\Omega_5^2 \right\} \ . 
\label{finite-c-sol}
\eeq 

Then, the supersymmetric solution is obtained as
\beq
e^\Phi= 1+\frac{q}{r^4}\ , \quad A(r)=1\, ,
\label{dilaton}
\eeq
where $M,~N=0\sim 9$ and
$R=\sqrt{\Lambda}/2=(4 \pi N_c)^{1/4}$. 
And $q$ represents the vacuum expectation value (VEV) 
of gauge fields condensate~\cite{Liu:1999fc,Ghoroku:2004sp}. 
In this configuration, the four dimensional boundary represents the 
$\cal{N}$=2 SYM theory. In this model, we find quark confinement in the
sense that we find a linear rising potential between quark and anti-quark
with the tension $\sqrt{q}/R^2$ \cite{Kehagias:1999iy,Ghoroku:2004sp}.
{However, chiral symmetry is preserved in the sense that the vacuum expectation
value of the order parameter is zero. In other words, the dynamical mass 
generation of massless quarks does not occur.}

\vspace{.3cm}
\section
{\bf Accelerating string solution}~

Here we restrict to the supersymmetric case.
The metric for the supersymmetric case (\ref{dilaton}) is given as, 
\beq\label{background-susy}
ds^2_{10}=e^{\Phi/2}R^2
\left\{
{u^2}\left(-dt^2+(dx^i)^2\right)+
\frac{1}{u^2} du^2+d\Omega_5^2 \right\} \ . 
\eeq 
where $u=r/R^2$. The world sheet coordinates of the string are taken as $(t,u)$,
and it is assumed to be stretching in the direction $x_1\equiv x=x(t,u)$. Then 
the induced metric on the world sheet of the string is given as
\bea
 g_{tt}&=&-e^{\Phi/2}R^2u^2(1-\dot{x}^2) \label{ws-metric1} \\
 g_{uu}&=&e^{\Phi/2}{R^2}\left({1\over u^2}+u^2{x'}^2\right) \label{ws-metric2} \\
 g_{ut}&=&g_{tu}=e^{\Phi/2}{R^2}u^2\dot{x}{x'} \label{ws-metric3} 
\eea

Then
the Nambu-Goto action for a string stretching in the $x$ direction is given as follows,
\beq
 S=-{R^2\over 2\pi\alpha'}\int dtdu~ e^{\Phi/2}\sqrt{1-\dot{x}^2+u^4{x'}^2}
\eeq
for the world sheet
of the string $(\tau,\sigma)=(t,u)$, where $r=R^2u$, $x'=\partial_u x$,
$\dot{x}=\partial_t x$ and
\beq
 e^{\Phi}=1+{\tilde{q}\over u^4}\, , \quad \tilde{q}={q\over R^8}\, .
\eeq
The equation of motion for $x(t,u)$ is obtained as
\beq\label{emo-susy}
 {1\over 2}{\Phi}'{u^4x'\over \tilde{g}^{1/2}}
  +\left({u^4x'\over \tilde{g}^{1/2}}\right)'
   -\partial_t\left({\dot{x}\over \tilde{g}^{1/2}}\right)=0\, ,
\eeq
\beq
   \tilde{g}=1-\dot{x}^2+u^4{x'}^2\, ,
\eeq
where dash and dot denotes the derivative with respective to $u$
and $t$ respectively as given above.

\vspace{.8cm}
\noindent{\bf $q=0$ case}

\vspace{.3cm}
Before solving (\ref{emo-susy}),
we briefly review Xiao's analytic solution given for $q=0$ or
$\Phi=0$, namely in $AdS_5$ background. In this case, the solution is
obtained as 
\beq\label{ads5}
 x(t,u)=\sqrt{t^2+f_0(u)}\, , \quad f_0(u)=1/a^2-1/u^2
\eeq
This solution
represents the accelerated quark in the supersymmetric Yang-Mills
theory in the deconfinement phase. The value $a$ denotes the acceleration
of the quark sitting at the boundary. The speed of the string to $x$-direction depends on 
the position $u$, and it exceeds light velocity in the region $u\leq a$.

We notice here the following point.
For this solution, the induced metric $g_{tt}$ of the string world sheet
is given as
\beq\label{ws-metric1-1}
 g_{tt}=-u^2{f_0(u)\over t^2+f_0(u)}\, .
\eeq
This implies that $g_{tt}$ has a zero point at $u=a$ and this point could be 
regarded as the horizon on the world sheet. So we could see the Hawking radiation
of fields on the world sheet. However, there is no horizon in the 
5D bulk background, so this situation is interpreted as the gauge field radiation
of the accelerated color charged particle, namely the quark.

In the present case, the world sheet metric of the accelerated string
are time-dependent and non-diagonal, then we should go to the other
coordinate which would provide a clear vacuum of the theory. This is performed
by the Rindler coordinate transformation given by 
(\ref{transform-1})-(\ref{transform-3})
in the next section. In this case, we find the following world sheet metric
\bea
 g_{\tau\tau}&=&-R^2(s^2-a^2) \label{ws-metric10} \\
 g_{ss}&=&R^2{1\over s^2-a^2} \label{ws-metric20} \\
 g_{s\tau}&=&g_{\tau s}=0 \label{ws-metric30} 
\eea
These represent the bulk metric at the same time, so we find the horizon of the
world sheet and the one of the bulk are common. As a result, we find the thermal
bath of the radiation in the new coordinate.
This is known as Unruh effect.

\vspace{.3cm}
Here we give some comment on the string.
In the new coordinate, the string world sheet is transformed to 
$(\tau,s)$, and it stretches in the
direction $\beta=\beta(\tau,s)$. 
Then the string solution given above is transformed as
\beq\label{sol00}
 \beta(s,\tau)=0\, .
\eeq
The new horizon point is equivalent to the point of $f(u)=0$, which is the
horizon point of (\ref{ws-metric1-1}). 
The string represented by (\ref{sol00}) is the straight line from the horizon
$s=a$ to the boundary $s=\infty$.
Then the part $u<a$ of the string in Minkowski coordinate has
been hidden in the thermal bath (inside the horizon) in the new vacuum.
We can see the similar situation for the accelerated solutions given in 
different form of background.

\vspace{.8cm}
\noindent{\bf $q>0$ case}

\vspace{.3cm}
In the next, we solve the above equation (\ref{emo-susy}) for the case of 
non-trivial dilaton, namely for $q>0$. Also in this case, we solve (\ref{emo-susy})
by assuming the following functional form,
\beq\label{susy-sol}
 x(t,u)=\sqrt{t^2+f(u)}\, .
\eeq
We would find $f(u)\simeq f_0(u)$ at large $u$ where the effect of $q$
could be neglected. But the function $f(u)$ deviates from $f_0(u)$
when $u$ decreases, then it may be written as
\beq
  f(u)=f_0(u)+\tilde{f}(u)\, ,
\eeq
where $\tilde{f}(u)$ is expected to be proportional to $q$. These points are
seen as follows.

\vspace{.3cm}
The equation (\ref{emo-susy}) is
rewritten as the one of $f(u)$ as follows,
\beq\label{emo-susy-2}
  \sqrt{f/{f'}^2+u^4/4}\left({u^4\over 2\sqrt{f/{f'}^2+u^4/4}}\right)'
   ={1\over f'}+q{u^3\over q+u^4}
\eeq
where prime denotes the derivative with respect to $u$.
We consider 
firstly the asymptotic
behavior near the boundary ($u\to \infty$). 
Substituting $f(u)=f_0(u)+\tilde{f}(u)$ into this equation
(\ref{emo-susy-2}), we find
\beq\label{tidf-l}
  \tilde{f}(u)={k\over u^3}+O(1/u^5)
\eeq
at large $u$. Here $k$ is an arbitrary constant as $a$ in $f_0$. The solution
has two arbitrary constants since the equation is the second order differential
equation. Then we can set the asymptotic value of $f(u)$ as $f(\infty)=1/a^2$
for any $q$ as the boundary condition used here. However this condition restricts
the boundary condition in the infrared region of small $u$.

\vspace{.3cm}
For the boundary condition, $f(\infty)=1/a^2$,
the solution with positive $q$ deviates definitely from $f_0(u)$ at small $u$.
In general the zero point of $f(u)$ increases with $q$, namely
\beq
   u_1>a \, , \quad {\rm for}~~~f(u_1)=0.
\eeq
This is obtained when we fix the point at $u=\infty$ as $f(\infty)=1/a^2$
for any $q$. On the other hand, the value of $f(\infty)$ moves to the large
value with increasing $q$ when we fix $u_1$. The situation depends on the 
boundary condition in solving the equation of motion of $f(u)$. 
The understandable situation would be to fix the acceleration of the quark as
$a$ by $f(\infty)=1/a^2$. In this case, the ``horizon'' $u_1$ on the string
moves to larger value {and the string
configuration $f(u)$ is modified from $f_0(u)$} with increasing $q$. However the Rindler temperature,
which is obtained after a coordinate transformation where the string
is seen to be static, {does not} depends on $q$ and also other parameter coming from 
the coordinate transformation.
{On the other hand, Rindler Temperature changes with the parameter
 of ERT, namely $a$.} We show this point in the next section.

\begin{figure}[htbp]
\vspace{.3cm}
\begin{center}
\includegraphics[width=7cm]{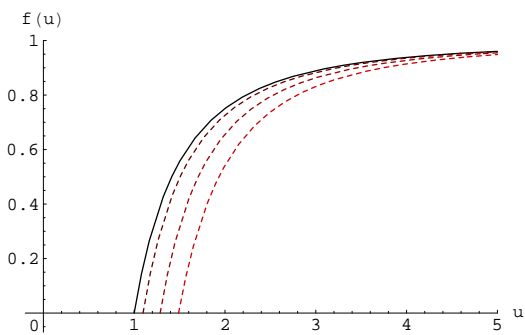}
\caption{{\small The numerical results of ${f}(u)$ for $a=1$ 
($f(\infty)=1/a^2=1$) and $q=0,~0.5,~3.0$ 
and $10$ are shown from left to the right.
The zero point of the solution moves to the right with increasing $q$.}}
\label{susy-f}
\end{center}
\end{figure}

For $q>0$, we give here the numerical 
solutions. The solutions for $q=0.5, 3.0, 10$ are shown in the Fig. \ref{susy-f} and
it is compared with the one of $q=0$. We can see the zero point of $f(u)$
moves to the larger value of $u$ as stated above.

\vspace{.3cm}
\section{Extended Rindler Transformation}
Then we move to the comoving coordinate of the accelerated
quark. Usually this is performed by the transformation among the two
coordinate in the 4D space-time, time and the one of the accelerated direction,
as seen in \cite{Unruh:1983ac,Paredes:2008cr}.
However here we use another form of
Rindler transformation proposed by Xiao \cite{Xiao:2008nr}.
This transformation is performed among the 
three coordinates, time, the accelerated direction
and the fifth coordinates of the original AdS$_5$ space-time, namely for $(t,x,u)$.
We call this as the extended Rindler transformation (ERT). After the transformation,
in the new coordinate, the accelerated quarks given above
in the original coordinate
are seen as static and they are in a thermal bath with a finite temperature.
Especially, we notice that these quarks are moving as free particles.
In this sense, the quark is not confined in the dual gauge theory for the
Rindler coordinate given by the extended Rindler transformation.
We study the details of this point in the followings. 
 
For the case of $q=0$,
the comoving coordinate, $(s,\tau,\beta)$,
of this accelerated quark is obtained from the original coordinates, $(u,t,x)$,
by the following extended coordinate
transformation \cite{Xiao:2008nr},
\bea\label{transform-0}
 x&=&\sqrt{{1\over a^2}-s^{-2}}~ e^{a\beta}\cosh(a\tau)\, , \label{transform-1}\\
 t&=&\sqrt{{1\over a^2}-s^{-2}}~ e^{a\beta}\sinh(a\tau)\, , \label{transform-2}\\
 u&=&s ~e^{-a\beta}\, \label{transform-3}
\eea
Then, the new metric is given as
\beq\label{transformed-metric}
 ds^2_{10}=R^2\left({ds^2\over s^2-a^2}-(s^2-a^2)d\tau^2
+s^2\left(d\beta^2+e^{-2a\beta}[dx_2^2+dx_3^2]\right)+d\Omega_5^2\right)
\eeq
and the above string configuration (\ref{ads5}) is given by
\beq
  \beta=0\,
\eeq
which is static since it is independent of the new time coordinate $\tau$.
{This string represents a free\footnote{Here, ``free'' means that the quark is not bounded with an anti-quark as a meson.} quark string which connects a D-brane put at
some $s>a$ and the event horizon $s=a$. }

We should notice here that we can replace the parameter $a$ by other value, for
example by $\tilde{a}$, in the above coordinate transformation (\ref{transform-1})-(\ref{transform-3}).
In this case, we find the transformed accelerated string at
\beq
  \beta={1\over \tilde{a}}\log(\tilde{a}/a)\,
\eeq
where we imposed the condition that the position of the quark on the boundary
$(x,t,u)=(1/a,0,\infty)$ is transformed
to $(\beta,\tau=0,s=\infty)$. 
{In this new coordinate with $\tilde{a}$,
the Rindler temperature is obtained as $\tilde{a}/2\pi$. 
We however
set here as $\tilde{a}=a$ for the simplicity. In this case, the Rindler temperature 
is directly related to the acceleration of the particles as in the particle
physics. }

\vspace{.3cm}
For the case of $q>0$, we consider the following similar transformation,
\bea
 x&=&g(s)~ e^{a\beta}\cosh(a\tau)\, , \label{trans} \\
 t&=&g(s)~ e^{a\beta}\sinh(a\tau)\, , \label{trans2}\\
 u&=&h(s) ~e^{-a\beta}\, \label{trans3}
\eea
where $a$ denotes the acceleration of the quark at the boundary.
Then we find the new coordinate,
\bea
 ds^2_{10}&=&e^{\Phi/2}\left\{
R^2h^2\left(\left({{h'}^2\over h^4}+{g'}^2\right)ds^2
-g^2a^2d\tau^2 \right. \right. \nonumber \\
 &&  \left. \left.+a^2\left(g^2+{1\over h^{2}}\right)d\beta^2
+e^{-2a\beta}\left[dx_2^2+dx_3^2\right] \right) +R^2d\Omega_5^2
\right\}\label{trans4}
\eea
where prime represents the derivative with respect to $s$, and we set as
\beq\label{relgh-1}
 (h^2)'=h^4(g^2)'
\eeq
to eliminate the $g_{s\beta}$.
Then $g$ is related to $h$ by solving
(\ref{relgh-1}) as
\beq\label{gh}
 g^2=c_0^2-{1\over h^2}
\eeq
where $c_0$ is an arbitrary constant.
Then we find again the Rindler coordinate for $c_0=1/a$ and $h=s$,
except for the prefactor $e^{\Phi/2}$.
In the present case, we change the radial coordinate from $s$ to
$h(s)$, then we have
\beq\label{Rindler-2}
 ds^2_{10}=e^{\Phi/2}R^2\left\{{dh^2\over h^2-a^2}-(h^2-a^2)d\tau^2
+h^2\left(d\beta^2+e^{-2a\beta}[dx_2^2+dx_3^2]\right)+d\Omega_5^2\right\}
\eeq
The accelerated strings therefore can be seen in the same 
Einstein frame coordinate also in the confining phase (for $q>0$).

\vspace{.3cm}
\noindent{\bf Transformed String Configurations}

\vspace{.3cm}
We consider how the string configuration of the accelerated strings
is seen in the comoving coordinate. 
The configuration is different from the $q=0$ case. 
The new configuration is shown in $\beta$-$h$ plane by using the relation,
\beq\label{Rindler-config}
 \tilde{f}(h~e^{-a\beta})={1\over a^2}\left(e^{2a\beta}-1\right)
\eeq
which is obtained from the form of the solution, $x^2-t^2=f(u)=f_0(u)+\tilde{f}(u)$.
Then, using the solution, $\tilde{f}(u)$, we find the string configurations
in the Rindler coordinate in terms of the equation (\ref{Rindler-config}).

\begin{figure}[htbp]
\vspace{.3cm}
\begin{center}
\includegraphics[width=7cm]{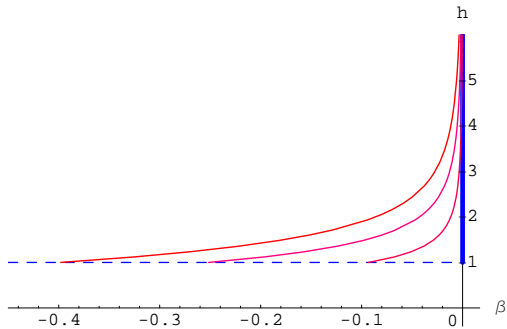}
\caption{{\small Examples of the string solution $\beta(h)$ for $q=0$ 
(straight line) and $q=0.5,3,10$ (from right to the left) with
$a=1$. 
The dotted line, $h=a(=1)$, represents
the Rindler horizon. Solutions for finite $q$ are bended due to the 
Yang Mills force expressed by the dilaton. Then
the larger $q$ becomes, the larger the deformation of the solution grows.}}
\label{beta-1}
\end{center}
\end{figure}

The boundary condition should be taken as $\tilde{f}(h=\infty)=0$ at fixed
$\beta$ since $a$ in $f_0=1/a^2-1/u^2$ represents the acceleration of the
quark at the boundary. This implies $\beta(h=\infty)=0$ from (\ref{Rindler-config}).
In the Fig. \ref{beta-1},
some examples of the solution for $q=10,~3,~0.5,~0$ with $a=1$ are shown.
When we set $q=0$, we obtain $\tilde{f}(u)=0$, which is the straight line
of $\beta=0$, namely the $h$-axis. The larger 
q becomes, the larger the deviation of the string configurations
from the straight line grows. (i) However, in any case, each 
string configuration
shows the free quark state in a heat bath of Rindler temperature
$a/2\pi$ and
(ii) the horizon is given by the point, $h=a$ (or $u=ae^{-a\beta}>a$), 
where the velocity of the
string to $x$-direction in the original coordinate arrives at the speed of the light.

\vspace{.6cm}
\noindent{\bf Temperature and Asymmetry of three space}

\vspace{.3cm}
Here we consider the temperature from two viewpoints. Firstly, it could be
given from the condition to evade a conical singularity in $(\tau,h)$ plane
for the Euclidean metric ($\tau\to i\tau$).

It is seen near the horizon.
By setting as $h=a(1+\epsilon^2/2)$,
the metric (\ref{Rindler-2}) is rewritten for small $\epsilon$ as follows
\beq
 ds^2_{10}=e^{\Phi_a/2}{R^2}\left\{d\epsilon^2
+a^2\epsilon^2d\tau^2 +\cdots\right\}\, ,
\eeq 
where 
\beq\label{dilaton-new}
 e^{\Phi_a}=1+{\tilde{q}~e^{4a\beta}\over a^4}\, .
\eeq 
From the above, the temperature is given by
\beq\label{R-temp}
 T_R={a\over 2\pi}\, .
\eeq
for fixed $\beta$.
Here we must notice that the prefactor $e^{\Phi_a/2}$ depends on the new
coordinate $\beta$, then the temperature 
is well defined only for fixed $\beta$ in this definition. 
However, we notice that the temperature given by (\ref{R-temp}) is
independent of $\beta$, 
in other words, the thermal equilibrium with the temperature $T_R$ would
be seen at any point of $\beta$, then this temperature 
is well defined in the three dimensional space,
$ds_{(3)}^2=d\beta^2+e^{-2a\beta}[dx_2^2+dx_3^2]$.

\vspace{.3cm}
The other way to define the temperature is given by considering 
the timelike Killing vector $\xi^{\mu}=\delta^{\mu 0}$, which satisfies
\beq
 \nabla_{\mu}\xi_{\nu}+\nabla_{\nu}\xi_{\mu}=0\, .
\eeq
In this case, the surface curvature at the horizon is given as
\beq
  k_H^2=-{1\over 2}(\nabla^{\mu}\xi^{\nu})(\nabla_{\mu}\xi_{\nu})
          =a^2\, ,
\eeq
then we obtain the same result with (\ref{R-temp}),
\beq
  T_H={k_H\over 2\pi |\xi^{\mu}|}={a\over 2\pi}=T_R\, .
\eeq

\vspace{.3cm}
So the temperature could be well defined in the Rindler coordinate, but
the three space, $ds_{(3)}^2=d\beta^2+e^{-2a\beta}[dx_2^2+dx_3^2]$,
is asymmetric. It is separated to the longitudinal ($\beta$)
and the transverse ($x_2$-$x_3$) directions. This asymmetric behavior 
 is also reflected to the color force, and it can be
seen through the dilaton which affects the force between the quark and the
anti-quark as being found through the Wilson-Loop, its analyses are given
in the next section. 

Actually, the gauge coupling constant is defined by $g^2_{\rm
eff}=e^{\Phi_a}$ and it depends on $\beta$ and $h$ as follows,
\beq\label{dilaton-h}
 g^2_{\rm eff}=1+{\tilde{q}~e^{4a\beta}\over h^4}\, .
\eeq 
This implies that
the Yang-Mills force depends on the energy scale $h$ and also on the coordinate
$\beta$ in the real three space. In the present case, the temperature is finite
and the Yang-Mills force between a quark and anti-quark is completely screened 
when they are separated by the distance ($L$) larger than a critical value 
($L_*$). Namely the quark is independent of the anti-quark which is separated
by the distance $L>L_*$, but we know that the quark can feel the force from the
anti-quark in the region of $L<L_*$ and this force is nearly equivalent to the
one given at zero-temperature.

In the present case, we find the linear rising force in the region of $L_0<L<L_*$,
where $L<L_0$ defines the ultra-violet region of the conformal symmetric limit. 
And we find the tension parameter \cite{Ghoroku:2004sp}
\beq
 \tau_{\rm eff}={\sqrt{q}\over 2\pi\alpha' R^2}
\eeq
at zero temperature in the present model. Then we expect the tension parameter
in the Rindler coordinate would be given by 
\beq\label{tension-2}
 \tau_{\rm R}={\sqrt{\tilde{q}}R^2~e^{2a\beta}\over 2\pi\alpha' }\, ,
\eeq
which is however coordinate dependent. We can assure this point through
the Wilson-Loop calculation given below.
Therefore we study the dynamical properties in this vacuum by separating to
two cases, the longitudinal and transverse directions in the three dimensional
space in the new coordinate.

\subsection{Wilson Loop and the Force Between Quarks}
\vspace{.3cm}
From the gauge coupling given above (\ref{dilaton-h}), we can say that
the force between the quark and anti-quark would
depend on $a$ and also on $\beta$.
In our original metric, the quarks are confined due to the strong infrared
gauge coupling constant, namely it diverges for $u\to 0$. In the 
new coordinate (Rindler coordinate),
the infrared strong-force would be
screened by the fluctuations of the thermal matter with the temperature
$T_R=a/(2\pi)$. The situation 
would be parallel to the case of the AdS-Schwartzschild
background which is dual to the high temperature gauge field theory. 
Due to this screening, we would find deconfinement phase in the 
Rindler vacuum. This is the Unruh effect in the confinement theory.
In order to assure this point, we study the force between the quark and 
the anti-quark, which are represented by the static strings in the
Rindler vacuum (\ref{Rindler-2}).

\vspace{.6cm}
\noindent{\bf (a) Strings Stretched to the longitudinal ($\beta$) direction}

\vspace{.3cm}
Consider the string which is extending to the $\beta$ direction. 
Taking its world sheet 
as $(\tau,h,\beta(h))$, then 
the Nambu-Goto action of this string is given as
\beq\label{ac-beta}
 S=-{1\over 2\pi\alpha'}\int d\tau dh R^2 e^{\Phi/2}
 \sqrt{1+h^2(h^2-a^2){\beta '}^2}
\eeq
where $\beta '=\partial_h\beta$, namely the prime denotes the 
derivative with respect to $h$.
The equation of motion for $\beta$ is given as
\beq\label{eq-beta}
  \partial_h\left({e^{\Phi/2}h^2(h^2-a^2){\beta '}\over
      \sqrt{1+h^2(h^2-a^2){\beta '}^2}}\right)=
      \sqrt{1+h^2(h^2-a^2){\beta '}^2}\partial_{\beta}e^{\Phi/2}
\eeq
We notice here that
the dilaton $\Phi$ depends on $\beta$ and $h$, and we find
\beq\label{eq2-beta}
  \partial_{\beta}e^{\Phi/2}=2a e^{-\Phi/2}{\tilde{q}e^{4a\beta}\over h^4}\, ,
\eeq
then the right hand side of (\ref{eq-beta}) is not zero, so we cannot obtain
the symmetric U shaped string configuration. We therefore solve the equation 
of motion by using a parameter $s$ along the string.
\footnote{The parameter $s$ introduced here has nothing to do with the one
given in (\ref{transform-1})-(\ref{transform-3}) in the above.}

\vspace{.3cm}
Before studying the Wilson loop, we like to show that the Rindler coordinate
is dual to the deconfinement phase of the gauge theory.
We consider the energy $E$ of the string, which 
is given by changing the variable from ${h,\beta(h)}$ to ${\beta,h(\beta)}$ as
\beq
    E={R^2\over 2\pi{\alpha'}}\int d\beta ~n\sqrt{1+{{h'}^2\over h(h^2-a^2)}}
\eeq
\beq
   n=e^{\Phi/2}h\sqrt{(h^2-a^2)}
\eeq
If quarks are confined, we would find
\beq
  E=\tau_{\rm eff}L\, , \quad  \tau_{\rm eff}={R^2\over 2\pi{\alpha'}}n(h^*)
\eeq
where $\tau_{\rm eff}$ denotes the tension of the linear rising potential
between the quark and the anti-quark, which are separated by $L$, and 
$h^*$ is the minimum point of $n(h)$. In the present case, $h^*=a$
and $n(a)=0$, so we cannot find linear potential with a definite tension
for large $L$. In other words, the quarks are
not confined. The main reason of this deconfinement might be the screening of the
color force at finite distance due to the thermal effect of the 
thermal bath with the Rindler temperature given above.

\vspace{.3cm}
\noindent{\bf Reparametrization invariant formulation}

These points can be seen more explicitly by calculating the Wilson loop
in the Rindler vacuum. As mentioned above, we rewrite the action (\ref{ac-beta})
by using the parameter $s$ along the string as follows
\beq\label{hamilton-beta1}
 S=-\int d\tau U\, ,
\eeq
\beq\label{hamilton-beta2}
 U={R^2\over 2\pi\alpha'}\int ds~ {\cal L}=\frac{R^2}{2\pi\a'}\int ds~ e^{\Phi/2}
 \sqrt{\dot{h}^2+h^2(h^2-a^2)\dot{\beta}^2}\, ,
\eeq
where the dot denotes the derivative with respect to $s$.
We notice that the above Lagrangian $U$ is reparametrization invariant
with respect to $s$, then we can give the Hamiltonian $H=H(h,p_h,\beta, p_{\beta})$
where 
\beq
  p_h={\partial {\cal L}\over \partial\dot{h}}\, , \quad 
  p_{\beta}={\partial {\cal L}\over \partial\dot{\beta}}\, , 
\eeq 
and we obtain
\beq
  H={\tilde{H}\over \Delta} \, , \quad \Delta^{-1}=2e^{-\Phi/2}
 \sqrt{\dot{h}^2+h^2(h^2-a^2)\dot{\beta}^2}\, , 
\eeq
\beq
  \tilde{H}={1\over 2}\left\{p_h^2+{p_{\beta}^2 \over h^2(h^2-a^2)}
      -e^{\Phi} \right \}\, .
\eeq
Notice that we can use the Hamiltonian $\tilde{H}$ instead of $H$. This is allowed since the theory we are solving is written in the
reparametrization invariant form. In this case, we impose the Hamiltonian
constraint as in the gravitational theory, 
\beq
  H=0\, \quad {\rm then} \quad \tilde{H}=0\, .
\eeq
Under this constraint, we find that the difference of the solutions of the Hamilton's
equations written by $H$ and $\tilde{H}$ is in the parametrization of the solution.
For example we write the solution as $h(s)$ or $h(f(s))$, where 
\beq
{\partial f(s)\over \partial s}=\Delta 
\eeq
So in both cases of $H$ and $\tilde{H}$, we will find the same string configuration.

Then we can derive the equations of motion of the string by the following
Hamilton's equations of $\tilde{H}$,
\beq\label{Hamilton-eq1}
  \dot{h}={p_{h}}\, , \quad
  \dot{\beta}={p_{\beta} \over h^2(h^2-a^2)}\, , 
\eeq
\beq\label{Hamilton-eq2}
  \dot{p_h}={2h^2-a^2\over h^3(h^2-a^2)^2}p_{\beta}^2
-{2\tilde{q}\over {h}^5}e^{4a\beta}\, ,\quad \dot{p_{\beta}}={2a\tilde{q}\over {h}^4}e^{4a\beta}\ .  
\eeq

\vspace{.3cm}
\noindent{\bf Boundary Condition and Numerical Solutions}
\begin{figure}[htbp]
\vspace{.3cm}
\begin{center}
\includegraphics[width=7cm]{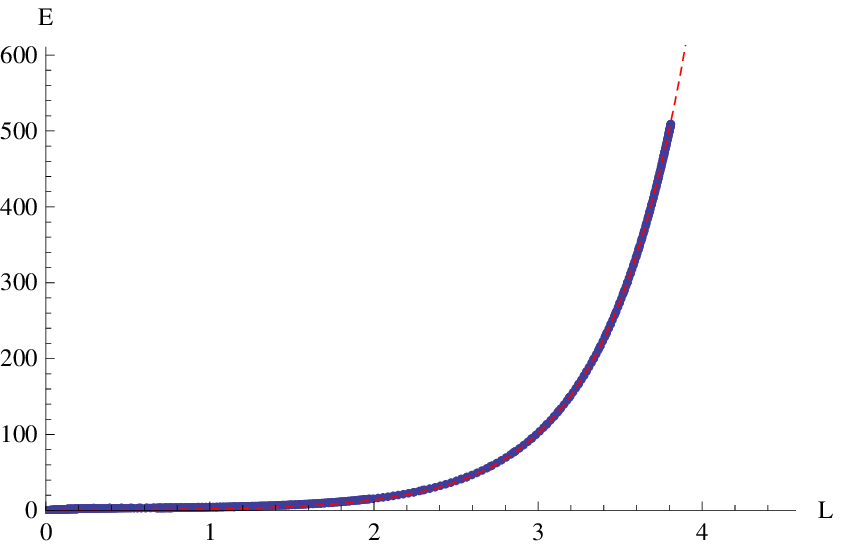}
\includegraphics[width=7cm]{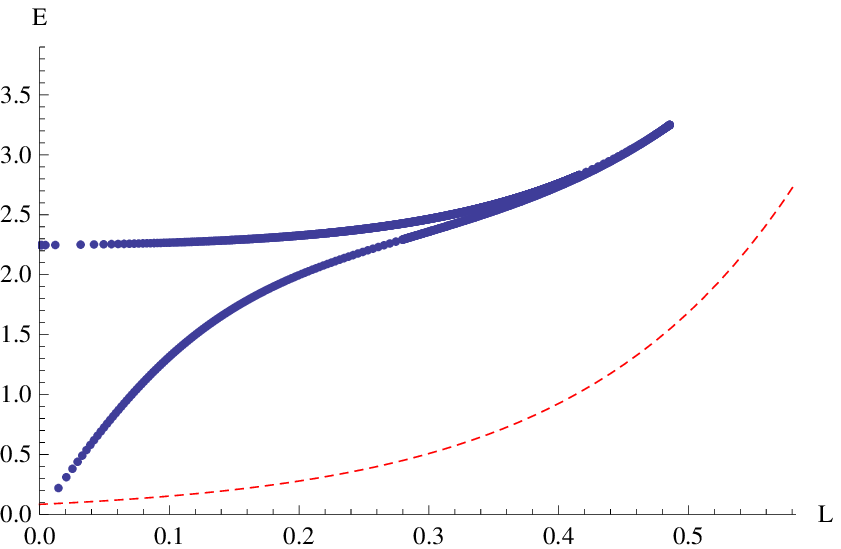}
\caption{{\small $E$-$L$ relations for quark and anti-quark in the longitudinally
 extended case for $h_{\rm max}=10$,
 $R=1$, $\tilde{q}=10$ and $a=1$(left), $a=3$(right). 
The each (red) dashed curves represents Eq.(\ref{tension-beta}), which
 is expected as the effect of the color force existed in the confinement
 phase.
}}
\label{EL-graph-1}
\end{center}
\end{figure}

\vspace{.3cm}
Solving the above equations (\ref{Hamilton-eq1}) and (\ref{Hamilton-eq2}) numerically, 
the solutions are obtained as the functions of $s$. We solve them with the 
boundary conditions,
\beq\label{boundary-1}
h_{max}>h(0)=h_0>a\, , \quad  \beta(0)=\beta_0\, , \quad p_h(0)=0
\eeq
and $p_{\beta}(0)$ is given from the constraint $\tilde{H}=0$ at $s=0$ as,
\beq
  p_{\beta}(0)=h_0\sqrt{(h_0^2-a^2)\left(1+{\tilde{q}e^{4a\beta_0}\over h_0^4}\right)}\, .
\eeq
Here $h_{max}$ denotes the end point of the string, and it is
fixed so that this point is interpreted
as the flavor brane's position or a UV cutoff nearby the boundary. 
And $h_0$ denotes the bottom of the
string solution. This point is varied to obtain the string solutions with
various different energy $E$, which is obtained by substituting the 
solution into the following equation,
\beq\label{energy-a}
  E={R^2\over 2\pi\alpha'}\int_{s_{dw}}^{s_{up}} ds~ {\cal L}\, .
\eeq 
Here the values of $s_{dw}$ and $(s_{dw}<)s_{up}$ are obtained from the 
solution $h(s)$ by solving the following equations,
\beq\label{s-max}
  h(s_{dw})=h_{max}=h(s_{up})\, .
\eeq
They indicates the two end points of the string solution.

As for $\beta_0$, it denotes the coordinate $\beta$ at the bottom of the
string solution and it controls the end point values $\beta(s_{dw})$ and
$\beta(s_{up})$. Then, we adjust $\beta_0$ such that $\beta(s_{dw})=0$
and $\beta(s_{up})\neq 0$. In this case, 
the distance ($L$) between the quark and anti-quark for these solutions is
given as
\beq
 L=|\beta(s_{up})-\beta(s_{dw})|=|\beta(s_{up})|\, .
\eeq
In our present calculation, $h$ is cut at $h_{max}(=10)$
and the quarks are supposed to be at this point. This regularization does not
affect on the large $L$ behavior of the Wilson loop estimations.

We show the results in the Fig.\ref{EL-graph-1}
for the region $\beta(s_{up})>0$,
where the parameters are set as
$R=1$, $q=10$, $a=1.0$(left) and $a=3.0$(right).

For sufficiently large $q$ compared to the temperature $a/2\pi$,
as mentioned above, we can see the color force with a definite
tension before the screening effects becomes dominant. 
In this region, the string energy is approximated as $E=\tau_{\rm R}L$
in terms of the tension $\tau_{\rm R}$ and the length $L$ of the string.
In the present case the tension is given by (\ref{tension-2}), however
$L$ is measured in the $\beta$ direction and the tension $\tau_{\rm R}$
depends on $\beta$. Therefore the energy could be estimated as follows,
\beq\label{tension-beta}
 E\simeq\int_0^{L}\tau_{\rm R}d{\beta}
       \simeq {\sqrt{\tilde{q}} R^2~e^{2aL}\over 4a\pi\alpha'}.
\eeq
Actually, we can see a good fit with this curve for low temperature case $a=1.0$,
and the numerical result as shown in the left figure of
Fig.\ref{EL-graph-1}. 

On the other hand, at high temperature, for the case of $a=3.0$,
we can find that the potential is screened before
it meets the dashed curve as shown in the right one of Fig.\ref{EL-graph-1}.
As for the screening,
we can see it 
through the existence of the
maximum point ($L_{\rm max}$) of $L$ which we discussed in previous
subsection. This point moves smaller $L$ with the increasing
temperature as expected from the usual high temperature theories
in the confinement phase.

The same analysis has been performed also for $\beta(s_{up})<0$. In this case,
the force becomes weak even if $a$ is small, so the screening becomes
dominant at rather small $L$ and it becomes difficult to see the remnant
of the confinement force. We abbreviated here to show the numerical results.

\vspace{.6cm}
\noindent{\bf (b) Strings Stretched to the transverse direction}

\vspace{.3cm}
Next, we consider the string stretched to the transverse direction, for example
to $x_2\equiv x$ direction which is transverse to the accelerated direction $\beta$.
In this case, the action is given as
\beq
    S=-{R^2\over 2\pi{\alpha'}}\int d\tau dh e^{\Phi/2}
   \sqrt{1+h^2(h^2-a^2){\beta'}^2+{x'}^2e^{-2a\beta} h^2(h^2-a^2)}
\eeq
where $x'=\partial x/\partial h$. And $\beta'=\partial \beta/\partial h$.
is also retained since we can't fix $\beta$ as a constant value as can be seen
from the equations of motion for $\beta$. 

In order to obtain the relation of $E$ and $L$ as given above, we must obtain
the string solutions. It is convenient to solve the equations after rewriting
the action in the reparametrization invariant form as above,
\beq
    U_{x}={R^2\over 2\pi{\alpha'}}\int ds {\cal L}_x
 ={R^2\over 2\pi{\alpha'}}\int ds e^{\Phi/2}
   \sqrt{\dot{h}^2+h^2(h^2-a^2)\dot{\beta}^2+\dot{x}^2e^{-2a\beta} h^2(h^2-a^2)}
\eeq
where dot denotes the derivative with respect to the parameter $s$. Then we have
the canonical momentum,
\beq
 p_h={\partial {\cal L}_x\over \partial\dot{h}}\, , \quad
 p_{\beta}={\partial {\cal L}_x\over \partial\dot{\beta}}\, , \quad
 p_x={\partial {\cal L}_x\over \partial\dot{x}}\, ,
\eeq
and the following Hamiltonian
\beq
  \tilde{H}_{x}={1\over 2}\left\{p_h^2+{p_{\beta}^2 \over h^2(h^2-a^2)}
      +{p_{x}^2 e^{2a\beta} \over h^2(h^2-a^2)}-e^{\Phi} \right \}
\eeq
Then we can solve the equations of motion by the following
Hamilton equations,
\beq\label{Hamilton-eq-x}
  \dot{h}={p_{h}}\, , \quad
  \dot{\beta}={p_{\beta} \over h^2(h^2-a^2)}\, , \quad
  \dot{x}={p_{x} e^{2a\beta} \over h^2(h^2-a^2)}\, , 
\eeq
\beq
  \dot{p_h}={2h^2-a^2\over h^3(h^2-a^2)^2}\left(p_{\beta}^2+ p_{x}^2 e^{2a\beta}\right)
-{2\tilde{q}\over {h}^5}e^{4a\beta}\, , 
\eeq
\beq\label{Hamilton-eq-x-2}
   \dot{p_{\beta}}=-{ap_{x}^2 e^{2a\beta}\over h^2(h^2-a^2)}
     +{2a\tilde{q}\over {h}^4}e^{4a\beta}\, , \quad \dot{p_x}=0.  
\eeq

\vspace{.3cm}
\noindent{\bf Boundary Condition and Numerical Solutions}

Solving the above equations numerically, we imposed the following
boundary conditions,
\bea
\beta(0)=\beta_0\, ,\quad h_{max}>h(0)=h_0>a\, ,\quad x(0)=0\, , \\
p_{\beta}(0)=0\, ,\quad p_h(0)=0\, ,
\eea
and $p_x(0)$ is given from the constraint $\tilde{H}_x=0$ at
$s=0$ as
\beq
  p_{x}(0)=h_0 e^{a\beta_0}
   \sqrt{(h_0^2-a^2)\left(1+{\tilde{q}e^{4a\beta_0}\over h_0^4}\right)}\, .
\eeq
\begin{figure}[htbp]
\vspace{.3cm}
\begin{center}
\includegraphics[width=8cm]{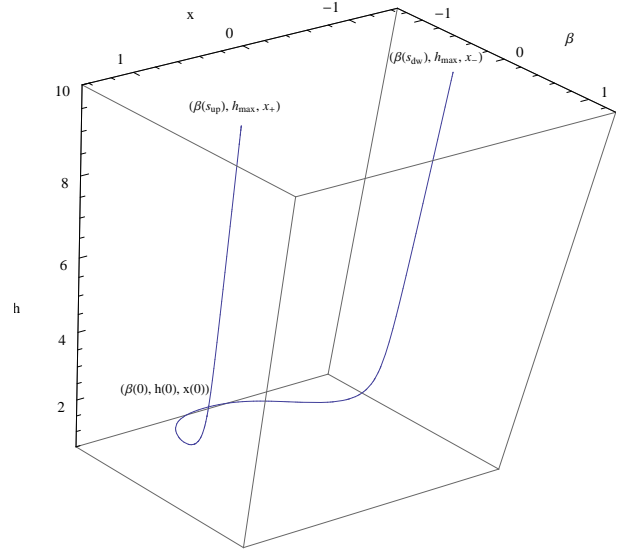}
\caption{{\small 3D string configuration stretched in the $y$-direction. As $y$
decreases the bottom of the string is stretched to $\beta$ direction.}}
\label{wil-susy-f}
\end{center}
\end{figure}

In this case, the distance between the quark and the anti-quark is measured
in the direction of $x$ by setting the coordinate $\beta$ as the same value at the two
end points.
Namely, we suppose at $h=h_{max}$,
\beq
x_+\equiv x(s_{up})\, ,\quad x_-\equiv x(s_{dw})\, , \quad 
    \beta(s_{up})=\beta(s_{dw})\equiv \beta_{\rm end},
\eeq
where $s_{up}$ and $s_{dw}$ is defined as (\ref{s-max}) and $x_+(s_{up})$ ($x_-(s_{dx})$) denotes the position of the quark
(anti-quark). Then the distance between quark and anti-quark is obtained as,
\beq
  L=x_+-x_-\, .
\eeq 

\vspace{.3cm}
We must be careful about the following fact that,
as shown above, the color force is in the present case depending on $\beta$
and $h$. Now we like to see the force in the $x$ direction through the solution
of (\ref{Hamilton-eq-x})-(\ref{Hamilton-eq-x-2}). So we should solve these equations
by imposing the condition that the end point coordinate $\beta=\beta_{\rm end}$
is kept as a fixed value.
So here we must tune the boundary values, $\beta_0$ and $h_0$ in order to 
realize the same $\beta_{\rm end}$ for each solution. Here, we obtain
the solution for $$\beta_{\rm end}=0\, ,$$ and
a typical string solution in the present case is shown in Fig.\ref{wil-susy-f}. 

\begin{figure}[htbp]
\vspace{.3cm}
\begin{center}
\includegraphics[width=6cm]{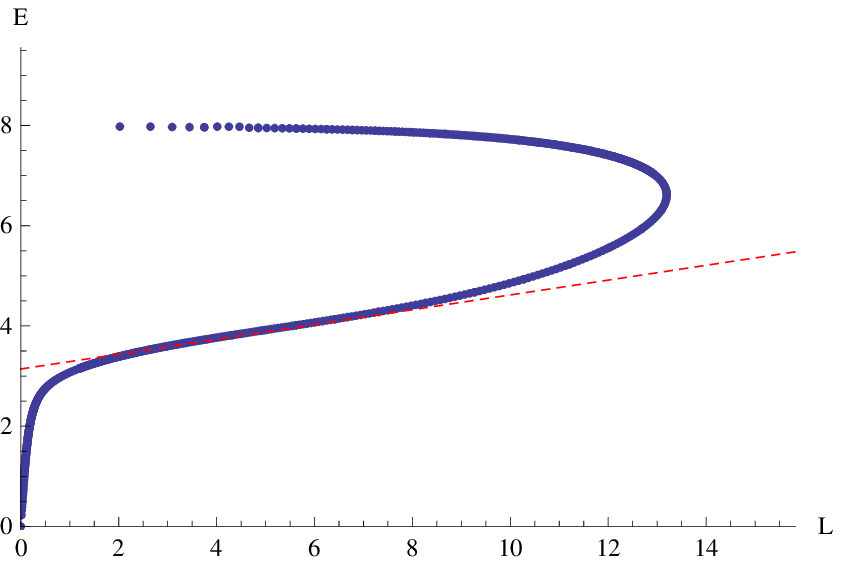}
\includegraphics[width=6cm]{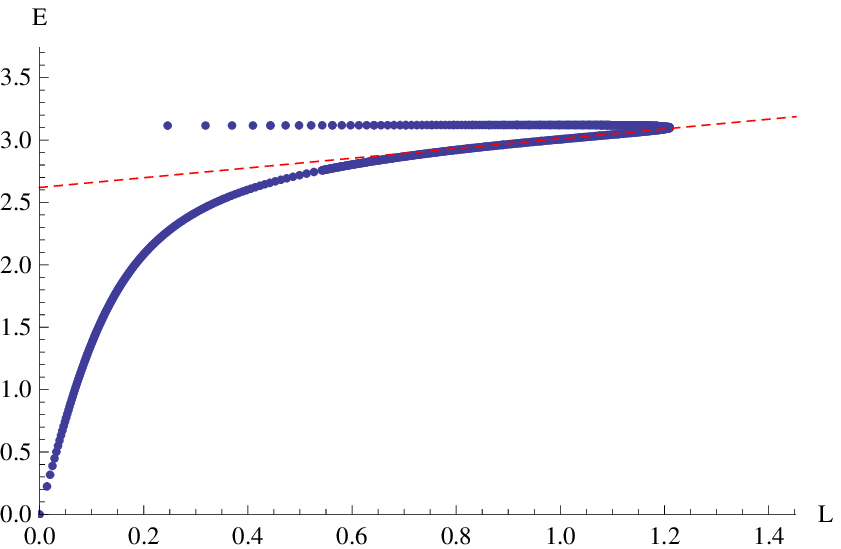}
\caption{{\small Typical $E$-$L$ relations for quark and anti-quark for
 $a=0.1$(left) and $a=1.0$(right), $h_{\rm max}=10$,
 $R=1$ and $\tilde{q}=10$. Each red dashed line represents the tension
 of the potential of linear rising part, and it is given by
 (\ref{EL-y1}) and (\ref{EL-y2}), which is expected as the effect of the
 color force existed in the confinement phase. 
}}
\label{EL-graph}
\end{center}
\end{figure}

Defining the energy of the quark and anti-quark system as (\ref{energy-a}) in 
the case of {\bf (a)},
$$
 E={R^2\over 2\pi\alpha'}\int_{s_{dw}}^{s_{up}} ds~ {\cal L}_x\, ,
$$ 
we can see the relation of $E$ and $L$ as above.
A typical results are shown in the Fig. \ref{EL-graph}. In this case we can see the linear rising part before
the screening. This rising part is fitted by the formula 
\beq\label{EL-y1}
  E={\sqrt{\tilde{q}}~e^{2a\bar{\beta}} R^2\over 2\pi\alpha'}L,
\eeq
where $\bar{\beta}$ would be approximately given by the $\beta_x$ on the horizon 
($h=a$) as,  
\beq\label{EL-y2}
  \bar{\beta} \approx\frac{\beta_0|_{h_{0}=a}}{3}.
\eeq

As for the upper part of $E$-$L$ relation, the curve increases with
decreasing $L$. This point is understood as follows. The upper part
is obtained by pulling the bottom point of the string to near the horizon.
Then the lower part of the string grows to the direction of $\beta$ and
the energy of the string becomes large as shown in Fig.\ref{wil-susy-f}.
This kind of behavior cannot be seen in the
case of the AdS-Schwartzschild background.

\vspace{.3cm}
\noindent{\bf Screening Length and Temperature}

\begin{figure}[htbp]
\vspace{.3cm}
\begin{center}
\includegraphics[width=7cm]{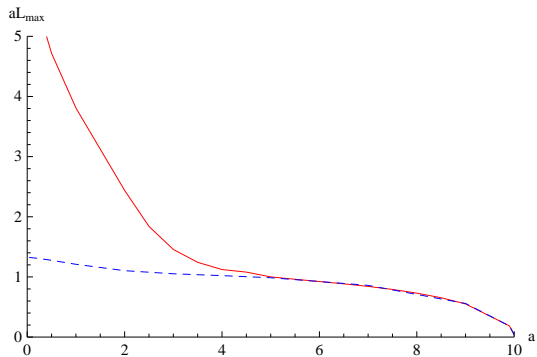}
\caption{{\small $a$-$aL_{max}$ relations for $\tilde{q}=10$, $R=1$ and
 $h_{max}=10(\geq a)$. The (Red) line is obtained from the string
 stretched to the longitudinal direction. The (Blue) dashed line is
 obtained from the string stretched to the transverse direction.}}
\label{a-Lmax}
\end{center}
\end{figure}

\vspace{.3cm}
In the next, we turn to the temperature dependence of the screening length, which
is defined by the maximum value of allowed $L$ for a given temperature. It is
denoted by $L_{\rm{max}}$, and it usually decreases with temperature as observed
in the theory dual to the AdS-Schwartzschild background \cite{Ghoroku:2005tf}.
We find finite $L_{\rm{max}}$ for any
finite $a$ or temperature $a/2\pi$ as the 
reflection of the screening of the color force.
Our result of the relation between $L_{max}$ and $a$ is
given in the Fig.\ref{a-Lmax}. 

The dotted (solid) curve represents the results from the Wilson loop stretched
in the transverse ($x$) (longitudinal ($\beta$)) direction. The behavior of
two curves clearly differs from each other in the small $a$ region. This 
implies that the color force enhancement in the longitudinal direction
at low temperature is greater than the force in the transverse direction.
In the case of the transverse direction, the behavior is similar to the case
of the AdS-Schwartzschild background, where $aL_{\rm max}$ is almost constant
or varies slowly with the temperature. On the other hand, at high temperature,
the screening becomes dominant and the asymmetric behavior disappears. As a result,
the two curves coincident at large $a$ (at high temperature) region.
This is also assured from the fact that the potentials in two cases 
approaches to the similar form. 

\vspace{.6cm}
\subsection{Trailing string and drag force}

Next we examine the drag force working on the
quark moving with a constant velocity in the hot gluons.
This is done by studying the trailing solution,
which was discussed in \cite{Herzog:2006gh,Gubser:2006bz,Gursoy:2009kk}
in the AdS-Schwartzschild background. In the present case,
it is performed in the Rindler background (\ref{Rindler-2}) given here. 
According to the work \cite{Herzog:2006gh,Gubser:2006bz,Gursoy:2009kk}, we consider a heavy quark
moving with a fixed velocity $v$ in the thermal medium. 
This running quark is expressed through a string with the velocity $v$, 
and its end point is on the boundary.

When we choose the coordinate $\beta$ as the moving direction, 
the string solution is supposed as
\beq\label{beta-t}
  \beta(\tau,h)=v\tau+\tilde{\xi}(h)\, .
\eeq
For the coordinate (\ref{Rindler-2}), 
however, it is easily found that there is no such a form of solution. The
reason is that
the dilaton depends on both time $\beta$ and $\tilde{\xi}(h)$ as
$e^{\Phi}=1+\tilde{q}e^{4a\beta}/h^4$. Then the equation of motion for
$\tilde{\xi}$ contains time explicitly through the dilaton as,
\beq
  e^{\Phi}=
          1+{\tilde{q}e^{4a(v\tau+\tilde{\xi}(h))}\over h^4}\, .
\eeq
As a result, $\tilde{\xi}$
should depends on $h$ and also on $\tau$ to satisfy the equation of motion.

\vspace{.3cm}
Therefore, the moving direction should be chosen
as $x_2$ or $x_3$ in order to obtain a string solution with a constant
velocity as given in the r.h.s. of (\ref{beta-t}) and to
see the conserved momentum flow along the string
from the boundary to the horizon. 
So we embed the string with its
world sheet ($\xi_1,\xi_2$) into the space $(\tau(=\xi_1),h(=\xi_2),\beta,x_2,x_3)$ 
through the ansatz that $x_3=$ constant and 
\bea
\beta=\beta(h)\, , \quad x_2\equiv y(\tau,h)=v\tau+\xi(h)\, .
\eea
We notice that we should keep $h$ dependence of $\beta(h)$ also in this case
due to the non-trivial dilaton. 
Then, the induced metric of the string is given as
\bea
 g_{\tau\tau}&=&-e^{\Phi/2}R^2\left((h^2-a^2)-h^2
         e^{-2a\beta}v^2\right) \label{trail-metric1} \\
 g_{hh}&=&e^{\Phi/2}R^2\left({1\over h^2-a^2}+h^2e^{-2a\beta}{\xi'}^2
      +h^2{\beta'}^2\right) \label{trail-metric2} \\
 g_{h\tau}&=&g_{\tau h}=e^{\Phi/2}R^2h^2e^{-2a\beta}v{\xi'} \label{trail-metric3} 
\eea
and the Nambu-Goto action of this string is written as
\bea
 S&=&\int d\tau dh {\cal L}_{\rm tr} \nonumber \\
  &=&-{R^2\over 2\pi\alpha'}\int d\tau dh 
  e^{\Phi \over 2} e^{-a\beta}  \nonumber \\
    & &\times\sqrt{
   \left(
    e^{2a\beta}-{h^2v^2 \over h^2-a^2}
   \right)(1+h^2(h^2-a^2){\beta'}^2)
   +h^2(h^2-a^2)\xi^{\prime 2}}\, .
\eea

\begin{figure}[htbp]
\vspace{.3cm}
\begin{center}
\includegraphics[width=5cm]{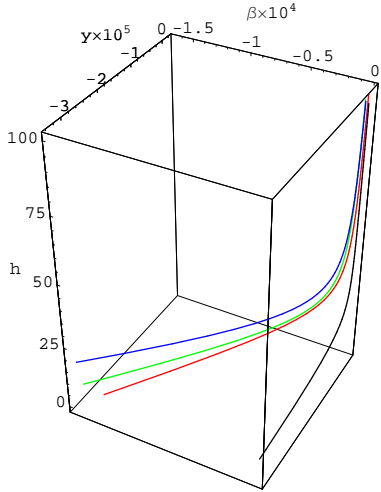}
\caption{{\small Strings trailing along with $y$-axis for $v=0.1$, $a=1$,
$R=1$ and $q=0$( on $\beta=0$ plane), $q=1$(front), $q=2$(middle), $q=5$(back).
These strings that have finite $q$  are curved to $\beta$-axis due to the existence of dilaton and the curve is sharper with the larger value of $q$.}}
\label{3D-drag-1}
\end{center}
\end{figure}

\vspace{.3cm}
From this action, we find equation of motion for $\xi$, 
and its conserved conjugate
momentum $\pi_{\xi}$ ($\pi_{\xi}'=0$) is obtained as
\beq\label{pixi}
  \pi_{\xi}={\partial {\cal L}_{\rm tr}\over \partial \xi'}=-{R^2\over 2\pi\alpha'}e^{\Phi \over 2} e^{-a\beta}
  {h^2(h^2-a^2)\xi^{\prime }\over \sqrt{
   \left(
    e^{2a\beta}-{h^2v^2 \over h^2-a^2}
   \right)(1+h^2(h^2-a^2){\beta'}^2)
   +h^2(h^2-a^2)\xi^{\prime 2}}}\, .
\eeq
Here we notice that we also find $\pi_{\beta}={\partial {\cal L}_{\rm tr}\over \partial \beta'}$. However there is no momentum in the $\beta$ direction now,
so we don't need to consider the drag force in this direction. However, we must
solve both $\beta(h)$ and $\xi(h)$ in order to obtain $\pi_{\xi}$ as seen from
Eq. (\ref{pixi}). As a result, the string configuration moving with a constant
velocity is deformed also to $\beta$ direction. The profile of such string
is shown in the Fig. \ref{3D-drag-1}.

\vspace{.3cm}
Here we give the following comment instead of showing the numerical
behavior of the drag force.
The drag force on the quark can be determined by the momentum flow 
$\pi_{\xi}$ which is
lost as the flow from the string to the horizon \cite{Gursoy:2009kk}. 
We find the horizon on the
string world sheet at $h=h_*$ from (\ref{trail-metric1}),
\beq
  h_*={a \over \sqrt{1-v^2e^{-2a\beta}}}\, ,
\eeq
which depends on $v$ and $\beta$. And $h_*>a$, namely, it is larger than 
the bulk horizon. Another point to be noticed is that the velocity is constrained
in the present case as
\beq
  v\leq e^{a\beta}
\eeq
for fixed $\beta$.

In any case, thus, we can estimate the drag force 
at the horizon $h=h_*$, where $\xi'$ and $\beta'$ dependence disappears, as
\beq
  F_{\rm drag}=\pi_{\xi}=-{1\over 2\pi\alpha'}\sqrt{
\left(h_*e^{-a\beta(h_*)}\right)^4R^4+{q \over R^4}
}~v\, .
\eeq
Here we notice the following two points. First, we can see
$\pi_{\beta}(h_*)=0$, then the force observed at the horizon $h=h_*$ 
is only the drag force $F_{\rm drag}$ for dragging the string to the 
$y$ direction. While the value of $h_*$ is determined by solving $\beta(h)$ 
for the first time, it is not performed here.
In the second, the force is modified by the color force
due to the dilaton from the form which is proportional to the temperature,
$F_{\rm drag}\propto T^2$. 

\vspace{.3cm}
Finally as for the friction constant $\eta$,
one can define it as
\beq
 F_{\rm drag}={dp\over dt}=-\eta p\, , \quad p=m_q{v\over \sqrt{1-v^2}}
\eeq
where $m_q$ denotes the quark mass, and $\eta$ is written as,
\beq
  \eta={\sqrt{1-v^2}\over 2\pi\alpha'm_q}\sqrt{
\left(h_*e^{-a\beta(h_*)}\right)^4R^4+{q \over R^4}
}
\eeq
For small $v\simeq 0$, we find $h_* \simeq a $ and 
$h_*e^{-a\beta(h_*)}\simeq  u_1$,
then we get 
\bea
   \eta\simeq {1 \over 2\pi \alpha' m_q} \sqrt{u_1^4R^4+{q \over R^4}}\, .
\label{eq:eta-dAdS-Unruh}
\eea
Then the friction constant is related to the radiation power given in 
$AdS_5$ background \cite{Xiao:2008nr}.

In the case of the $AdS_5$-Schwarzschild background, 
the friction constant 
is given as follows \cite{Gubser:2006bz},
\bea
\eta_{AdS} = {\pi \over 2\alpha' m_q} R^2T^2.
\label{eq:eta-AdS-Sch}
\eea
On the other hand, in the limit of $q=0$, (\ref{eq:eta-dAdS-Unruh}) leads to
\bea
\eta \simeq {2\pi \over \alpha'  m_q}R^2T^2,
\label{eq:eta-AdS-Unruh}
\eea
since $u_1=a$ in this limit.
Thus we recognize the difference of factor four between
(\ref{eq:eta-AdS-Unruh}) and 
(\ref{eq:eta-AdS-Sch}). 
This point should be explained from some physical insight, but it is 
remained as an open problem here.

\vspace{.3cm}
\section{Relation to the 4D Field Theory}

Here we give a comment on the statement for the Unruh effect given
in \cite{Unruh:1983ac}, where the analysis is performed within the 4D
field theory. The main result 
is that the vacuum expectation values (VEV) of any Green's functions in the
vacuum of Minkowski space-time are the same with the one of the Rindler space-time
when the calculation is restricted to the same Rindler wedge in the
Minkowski coordinate. So one may consider
that the phase of the Minkowski vacuum cannot be changed in the
Rindler vacuum since the VEV of any order parameter would be the same in both vacuum.

However, in the present paper, we show that the vacuum of the Rindler space-time
is in the quark deconfinement phase in spite of the fact that the original
theory in the Minkowski vacuum is in the confinement phase.
Then our calculation seems to be inconsistent with the statement of
\cite{Unruh:1983ac}, however
this point would be resolved as follows.

\vspace{.3cm}
Since the confinement or deconfinement is discriminated by the VEV of the
Wilson loop, we concentrate on this quantity here.
In the field theory side, the corresponding operator would be given as, 
\beq
 {\cal O}={\rm tr}\left({\cal P}e^{ig\oint_{\cal C} A_{\mu}(z)dz^{\mu}}\right)
\eeq
where ${\cal P}$ denotes the path ordering of a closed path ${\cal C}$ in the
line integration for the gauge field $A_{\mu}(z)$.
Its vev is 
written as
\beq\label{Min-vev}
 A=\langle 0|{\cal O}|0 \rangle
\eeq
for the Minkowski vacuum $|0 \rangle$, and 
\beq\label{Rin-vev}
 B={{\rm Tr}\left(e^{-H^R/T}
{\cal O}\right)\over {\rm Tr}e^{-H^R/T}}
\eeq
for the finite temperature ($T$)
Rindler vacuum respectively. The statement in \cite{Unruh:1983ac} implies
the equivalence of $A$ and $B$ 
when they are
calculated within the same Rindler wedge. 

\vspace{.3cm}
In our holographic approach, 
the Wilson-loop defined above is estimated
in terms of the (static) string configurations,
whose end point are on the path ${\cal C}$, for both $A$ and $B$. 
The string configurations
are obtained as the classical solutions of the Nambu-Goto action embedded
in each background. 
Now, we perform the calculation for $A$ for a fixed path ${\cal C}$, then 
the calculation for $B$ has been 
done by using the path and the string configurations, which are
all obtained by ERT from those used in $A$. In this case, we will find
$A=B$. However, we didn't do the calculation in this way. 

Our calculation of $A$ and $B$ does not leads to $A=B$ due to the 
following three reasons. 
In the first, the paths used in $A$ and $B$ are not related by the ERT.
In order to see the potential between the quark and the anti-quark,
we have performed the calculations for 
the rectangular path in the $t$-$x$ plane 
for $A$ and the one in the $\tau$-$\beta$ plane for $B$ respectively. 
In this case, the rectangular path 
used in $A$ cannot transformed to the one used in $B$ 
by the ERT since it must be transformed to the $\tau$ dependent path. 

Secondly, the static solutions used in the evaluation of 
$A$ cannot be transformed to 
the static ($\tau$-independent) one 
used in $B$ since the static solutions given in the
Minkowski vacuum are generally transformed by ERT
to the $\tau$-dependent one.
Then our Wilson-loop calculations in 
$A$ and $B$ are not related by the ERT. We derived our result from them.
Actually,
we could obtain different 
results from the calculation of $A$ and $B$, the linear confinement potential
for $A$ and the screening and the deconfinement for $B$ respectively.

Thirdly, we should give a comment for the quark-string configuration
in the Rindler vacuum. This point is also related to the fact that the quark
in the Rindler vacuum is different from the one in the Minkowski vacuum.
In estimating $B$, the essential string solution responsible for the
proof of the deconfinement 
is the one which connects the boundary
and the event horizon, 
because this solution can be interpreted as the free quark and this is
possible only for the deconfinement phase.
We could find such a solution only in the Rindler vacuum. 
The interesting point 
is that this free quark configuration in the Rindler
vacuum is obtained by the ERT 
from the constantly accelerating quark-string configuration 
given in the Minkowski vacuum as shown above. However this configuration
is not used in the evaluation of $A$
in our theoretical scheme.
Due to these reasons, we are not seeing the relation $A=B$. We are examining
the parts, which cannot be related by ERT, of $A$ and $B$, 
then our statement is not contradicting with the one in \cite{Unruh:1983ac}.

\vspace{.3cm}
Of course, there are remaining problems related to the coordinate transformation
which is adopted in the present paper. We should study 
the other form of the coordinate transformation. For example, 
the coordinate transformation given in \cite{Unruh:1983ac}
can be considered as such a transformation, which does not include the fifth
coordinate of the bulk. This
is different from the ERT used here. In the latter case,
the transformation is performed in three dimensional coordinate including the
fifth one. As a result, the event horizon appears in the bulk, 
then the infrared region is cut off in the dynamics of the dual
4D theory. Then the dynamical properties responsible to the long range force
would be lost in the vacuum of the new coordinate. 

We should study also the VEV of other physical quantities.
In this context, we notice another work given in \cite{Ohsaku:2004rv} from
other 4D non-perturbative approach, and the author shows chiral symmetry
restoration in the Rindler vacuum. 
So it would be necessary
to proceed the work more in this direction in order to make clear
the Rindler vacuum.

\vspace{.3cm}
\section{Summary and Discussions}
We give here a
constantly accelerated quark as a string solution of the 
Nambu-Goto action which is embedded in the supergravity background dual to the
confining Yang-Mills theory. For this accelerated quark given  
in the zero-temperature Minkowski space-time, 
we find an event horizon in its world sheet metric. 
This horizon is also found in the case of AdS$_5$ background
dual to the non-confining theory. In any case,
this fact can be
considered as a clear signal of the radiation of
gluons due to the acceleration of the color charged quark.

We consider the extended Rindler transformation proposed by Xiao
in order to move to the comoving frame of the accelerated quark and to study
the properties of the Rindler vacuum.
The coordinate transformation is generalized to the 5D bulk theory, but
the boundary is described by the usual 4D Rindler metric.
In this case, the dual theory is found in a thermal
medium with the Rindler temperature, so it is expected that the theory is in a
different vacuum from the one in the inertial frame. To study this
point, 
we have examined several dynamical properties of the new vacuum to
compare them to the one observed in the inertial coordinate. 

We could find 
that the vacuum properties are changed from the one seen in the inertial frame.
In the Rindler vacuum, the color force
is screened by the thermal effect at long distance. As a result,
the confining property has been lost.
The screening length depends on the temperature and also on the direction
in the three space. As for the temperature dependence, we find some 
difference from the one observed previously
in the AdS-Schwartzshild background, especially for the screening in the
longitudinal direction.
The reason of the anisotropy in the
3D space, longitudinal and transverse to the acceleration, can be 
reduced to the behavior of the dilaton. The dilaton expresses the
gauge coupling constant, and it is deformed in the longitudinal direction.
due to the extended Rindler transformation. 
Then the potential between the quark and anti-quark has different form 
in the longitudinal and the transverse directions.
This point has been assured through the direct observation
of the potential between the quark and anti-quark.

Even if we calculate the force in the transverse direction, we can see the
effect of the longitudinally deformed force.
For example, consider some excited bound state of the quark and anti-quark,
then we could find a small repulsion in the
direction ($y$) transverse to the accelerated direction ($\beta$). 
This repulsion is understood 
from the string configuration in the 3D space $\beta$-$y$-$h$. When the distance $y$
decreases, the string is stretched in the $\beta$ direction near the horizon
$h\simeq a$. This phenomenon
is understood as
the remnant of the strong confining force near the horizon.
On the other hand, the 
attractive force is exponentially enhanced in the longitudinal direction at fairly
large distance. This behavior can be reduced to the deformed dilaton in Rindler
coordinate in the longitudinal direction.

We also discuss the drag force to see the thermal effects in the Rindler vacuum. 
We find the friction constant is related to the radiation power of the accelerated quark in the original vacuum.
And we recognize the difference of factor four between the friction constant
in the Rindler vacuum and the one in the $AdS_5$-Schwarzschild background.

\vspace{.3cm}
Finally, we give our main conclusion that we could find a new vacuum when we 
move to the comoving coordinate of an accelerating 
quark by the extended Rindler transformation considered here
and find a kind of a high temperature theory of deconfinement phase.
Then, in the new vacuum, the properties given by the long range force property
in the inertial frame have been lost. Thus our calculation seems to be 
inconsistent with the claim given in \cite{Unruh:1983ac}, however
this point is resolved as explained in the previous section. The main reason
is that we are considering the different VEV of Green's functions to decide the
phase of the vacuum.

\vspace{.3cm}
\section*{Acknowledgments}



\newpage
\end{document}